# Irreversible reorganization in a supercooled liquid originates from localised soft modes

Asaph Widmer-Cooper[1,*], Heidi Perry[2,*], Peter Harrowell[1,#], David R.Reichman[2,#]

[1] *School of Chemistry, University of Sydney, Sydney, New South Wales 2006, Australia*
[2] *Department of Chemistry, Columbia University, 3000 Broadway, NY, NY, 10027, USA*

**The transition of a fluid to a rigid glass upon cooling is a common route of transformation from liquid to solid that embodies the most poorly understood features of both phases[1,2,3]. From the liquid perspective, the puzzle is to understand stress relaxation in the disordered state. From the perspective of solids, the challenge is to extend our description of structure and its mechanical consequences to materials without long range order. Using computer simulations, we show that the localized low frequency normal modes of a configuration in a supercooled liquid are causally correlated to the irreversible structural reorganization of the particles within that configuration. We also demonstrate that the spatial distribution of these soft local modes can persist in spite of significant particle reorganization. The consequence of these two results is that it is now feasible to construct a theory of relaxation length scales in glass-forming liquids without recourse to dynamics and to explicitly relate molecular properties to their collective relaxation.**

A crucial concept[4-9] for the transition between liquid and glass is that the dramatic increase in viscosity of a supercooled liquid as it approaches vitrification is caused by the growth of localized domains of particles that must rearrange for the liquid to flow. Locating a causal link between local structure and such dynamical heterogeneities has proven elusive[10,11]. Recently, a fruitful computational approach to isolating the structural origin of dynamical heterogeneity has been put forward in the dual notions of an "iso-configurational ensemble" and "propensity for motion" [12]. The iso-configurational ensemble refers to the ensemble of trajectories that are run from an identical configuration of particles with random initial momenta sampled from the equilibrium Boltzmann distribution. Propensity refers to the mean squared displacement of individual particles when averaged over the ensemble at a given time scale. The heterogeneous character and increased clustering exhibited in the spatial propensity maps established that the spatial distribution of these dynamic heterogeneities can be explicitly attributed to structural features, as yet unidentified.

Our goal is to understand that aspect of dynamic heterogeneity directly associated with structural relaxation. To this end we are interested in motions that a) involve local reorganisation of particle configurations (here measured in terms of changes in nearest neighbour pairings[13]) and b) are irreversible over some observation time[14] and, hence, contribute to relaxation. To investigate this we consider a two dimensional binary mixture of soft discs whose transition from liquid to glass-like behaviour has been well characterised[15]. The temperature unit and the time scale $\tau$ are defined in the Methods Summary. We first establish how many nearest neighbours must be lost around a given particle before the probability of the tagged particle recovering its original environment falls below 5%. For the binary mixture of soft discs under investigation here, this threshold is four neighbours (see Supplementary Information). Equipped with this measure, we can determine how the *irreversible reorganisation* (IR) is distributed in a given configuration. We record, over an ensemble of 100 iso-configurational runs, the number of runs in which each particle meets the irreversibility criterion within a time interval of $200\tau$ (this corresponds to the time at which the peak of the non-Gaussian parameter[9] occurs at this temperature and represents a time scale of about 2000 collision events). Maps of the log of this probability distribution for six configurations at T = 0.4 are shown in Fig. 1. The irreversible reorganization mapped in Fig.1 are elementary components of the slow structural relaxation characterised by a time $\tau_\alpha$ (= $673\tau$ for the simulations reported here[15]).

What aspect of the initial configuration is responsible for the observed spatial distribution of irreversible events? Previously, we have demonstrated that the spatial distribution of the free volume[11] and local potential energy[10] did not exhibit any significant correlation with the spatial distribution of the propensity. To move beyond these purely local measures, we have determined the normal modes for the local potential energy minimum (the 'inherent structure') associated with a number of initial configurations. We shall refer to these as quenched normal modes. The participation fraction of particle $i$ in eigenmode $\mathbf{e}_\omega$ is given by $p_i = |\mathbf{e}^i_\omega|^2$. In Fig. 2 we have plotted maps of the particle participation fractions summed over the 30 lowest frequency modes for each of our initial configurations. An examination of the individual mode eigenfunctions (see Supplementary Information) indicates that these low frequency modes include both localised and delocalised character[18-20].

To establish the connection between the mode structure and the subsequent irreversible reorganization, we have plotted in Fig. 3 the positions (white circles) of those particles with $\geq 0.01$ probability of meeting our IR condition during the entire $200\tau$ interval in the iso-configurational ensemble on top of the maps of the participation fractions for the low frequency modes at time t = 0. Note that the majority contribution to the IR map comes from particles that lost their fourth neighbour late in the trajectory. While these results do not address the question of *when* or even *if* a given soft local mode will become involved in IR, they do strongly support a picture in which the irreversible reorganisation of a configuration originates from these modes.

Two points are worth emphasising. The mode participation fractions, whose spatial distributions are mapped in Fig. 2, are properties of the *static* initial configurations. Our demonstration of a strong correlation between the mode maps and the irreversible reorganization maps constitutes a significant success in understanding how structure determines relaxation in an amorphous material. Indeed, as Fig.2 illustrates, one may provide semi-quantitative prediction of IR domains as they emerge at relatively long times from the initial configuration alone. The second point is that, since we have used *quenched* modes, we have only used information about the *bottom* of the local potential minima. While it is quite possible that the time scale required for a reorganisation event will depend on the energy barriers associated with the transition, our results indicate that the spatial structure of such events is largely determined by the distribution of soft quasi-localised modes in the initial configuration.

Given our conclusion that relaxation originates with soft quasi-localised modes, it follows that our capacity to predict the subsequent spatial distribution of the irreversible relaxation depends on how persistent the mode distribution is in a configuration. After all, should the mode maps evolve rapidly then the structural information in a given configuration would quickly become irrelevant. The fact that we observe strong spatial correlations between the initial modes and relaxation some $200\tau$ later indicates that the spatial structure of the modes does generally persist over such times. This is remarkable given that small variations in the quenched modes, indicative of a change in the local minimum (or inherent structure), occur over $\sim 1\tau$ intervals. Details of these rapid changes are provided in the Supplementary Information. We conclude that the spatial structure of the quenched soft modes can often persist over many changes in the inherent structure. Preliminary results indicate that this persistence is also found in 3D mixtures (including temperatures below the empirical mode coupling temperature). [Supplementary Information].

We do, however, see examples where the mode structure is not so stable. An example of this is shown in Fig. 4. In Figs. 4a and 4b we compare the mode participation map for the initial configuration with the map of the maximum participation fraction observed per particle over five $10\tau$ runs starting from the configuration in Fig. 4a. The difference in spatial structure between these maps is a measure of the degree of variability of the mode structure. In Figs. 4c and 4d we overlay the particles exhibiting IR within $200\tau$ (as defined in Fig. 3) over the maps of Figs. 4a and 4b, respectively. While the mode structure of the initial configuration does not provide a quantitative predictor of the spatial distribution of IR (see Fig. 4c), the cumulative mode structure sampled over the multiple short runs does (see Fig.4d). This result demonstrates that even when the soft mode structure is not stable, the irreversible relaxation still originates with these soft modes, only now this IR is not well predicted by any single configuration. It appears that configurations such as that analysed in Fig.4 represent those caught in transit between configurations with more stable mode structure.

In this paper we have presented two important results relating to the slow relaxation in a model supercooled liquid. The first is that the irreversible reorganisation originates at the sites of the low frequency quasi-localised quenched modes. The second is that these modes typically persist for time scales significantly longer than the lifetime of a given inherent structure. These results show that the spatial location and extent of IR regions at relatively long times may be reasonably predicted by a simple, static property of the static initial condition. A number of previous reports have linked localised dynamics with the presence of soft modes[16-20] during or immediately prior to the appearance of the motion in question. In contrast, we have demonstrated that soft localised modes are typically present in configurations well in advance of the appearance of the irreversible reorganization associated with them. The results of this paper suggest that the quasi-localised modes can provide a unification of many of the major themes in current research on the glass transition. Can the growth in the four point susceptibility $\chi_4$ near the glass transition[6,21] and the jamming transition in granular material[22], along with the growth in the related kinetic correlation length, be directly connected to an equilibrium correlation function of the soft modes distribution? A recent paper[22] by Zeravcic et al has examined the localisation of vibrational modes in granular material. Does the persistence of the mode structure in real space reflect the transient confinement of the system trajectory within a 'metabasin' in configuration space[24]? How both the spatial character of the quasi-localized normal modes and their correlation with subsequent relaxation varies with temperature remains the most fundamental question raised by this paper and is the subject of ongoing research. We have established that these quasi-localised modes represent the strongest causal link yet established between structure and dynamic heterogeneity and, hence, an exciting route forward to establish how molecular properties influence relaxation in the supercooled liquid.

27. We would like to thank L. Berthier, G. Biroli, J. P. Bouchaud, A. Heuer, and C. O'Hern for useful discussions. HP and DRR would like to thank P. Verrocchio for providing the equilibrated 3D configurations NSF for financial support. AW and PH acknowledge the support of the Australian Research Council.

(*)-These authors contributed equally to this work. (#)-Authors to whom correspondence should be addressed: P. Harrowell(peter@chem.usyd.edu.au), D.R.Reichman(reichman@chem.columbia.edu)

**Materials**

For a glass-forming liquid, we use a two-dimensional (2D) equimolar binary mixture of particles interacting via purely repulsive potentials of the form

$$u_{ab}(r) = \varepsilon \left[\frac{\sigma_{ab}}{r}\right]^{12} \qquad (1)$$

where $\sigma_{12} = 1.2 \times \sigma_{11}$ and $\sigma_{22} = 1.4 \times \sigma_{11}$. All units quoted will be reduced so that $\sigma_{11} = \varepsilon = m = 1.0$ where m is the mass of both types of particle. Specifically, the reduced unit of time is given by $\tau = \sigma_{11}(m/\varepsilon)^{1/2}$. The average collision time at T=0.4 in the binary mixture is $0.1\tau$. The reduced unit of temperature is $k_B/\varepsilon$. A total of $N = 1024$ particles were enclosed in a square box with periodic boundary conditions. Molecular dynamics simulations were performed in the NPT ensemble using a Nosé-Poincaré-Andersen algorithm developed by Sturgeon and Laird[25]. The structural or 'alpha' relaxation time $\tau_\alpha$ is defined as the time required for the self intermediate scattering function $F_s(q,t)$,

$$F_s(q,t) = \frac{1}{N}\left\langle \sum_{j=1}^{N} \exp\{i\vec{q}\cdot[\vec{r}_j(0)-\vec{r}_j(t)]\}\right\rangle \qquad (2)$$

to decay to a value of $1/e$. The magnitude of the wavevector $q$ is set equal to the value at the first Bragg peak.

For the normal mode analysis, the inherent structure of each configuration was found using the conjugate gradient method. The dynamical matrix of the inherent structure was then defined, $\mathbf{D} = \dfrac{\partial^2 \Phi(r)}{\partial r_i^k \partial r_j^k}$ where $r_i^k$ is the $k^{th}$ component of the position $\mathbf{r}_i$ of particle $i$, $\Phi(r) = \sum_{i=0}^{N}\sum_{j=1,j\neq i}^{N} u_{ab}(r_{ij})$, $u_{ab}(r)$ is the intermolecular potential from Eq.1 and $r_{ij} = |\mathbf{r}_i - \mathbf{r}_j|$. The dynamical matrix was diagonalized using the template numerical toolkit[26].

To visualize the spatial distribution of the propensity, it is useful to remove the additional complexity of the configuration and use contour plots. As the data points are located at irregularly spaced particle coordinates, it is necessary to interpolate between them. We have used the modified version of Shepard's method as implemented in the NAG libraries.

**Figure Captions**

**Figure 1.** Contour plots of the probability ($\log_{10}$) of a particle losing 4 original neighbours, the criterion for irreversible reorganization, over 100 iso-configurational runs for 6 different initial configurations.

**Figure 2.** Contour plots of the participation fraction summed over the 30 lowest frequency modes for the quenched initial configurations of the same 6 configurations used in Fig. 1.

**Figure 3.** Contour plots of the low frequency mode participation (as in Fig. 2), overlaid with the location of particles (white circles) whose iso-configurational probability of losing 4 initial nearest neighbours within $200\tau$ is greater than or equal to 0.01.

**Figure 4.** a) Contour plot of the participation fraction summed over the 30 lowest frequency modes for a quenched configuration. b) Contour plot of the maximum value of the participation fraction observed over five $10\tau$ runs starting from the configuration in Fig. 4a. c) Particles whose iso-configurational probability of losing 4 initial nearest neighbours within $200\tau$ is greater than or equal to 0.01 (white circles) overlaid on the participation fraction map for the initial configuration. d) As in c) except that the overlay is over the map of the maximum participation fraction shown in b).

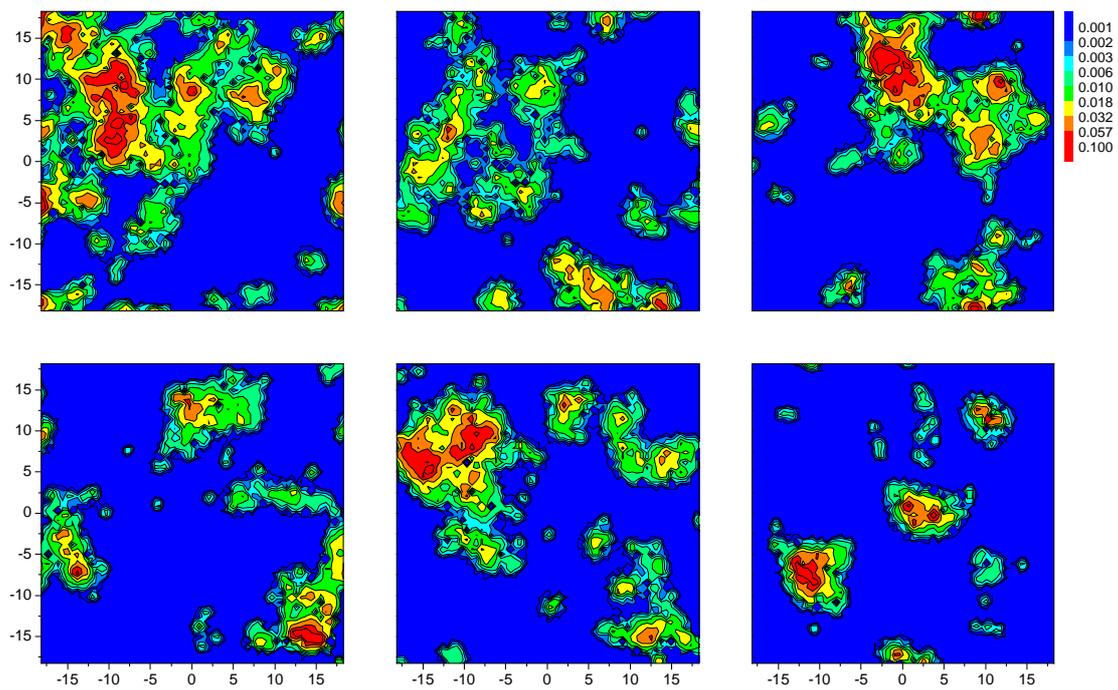

Figure 1

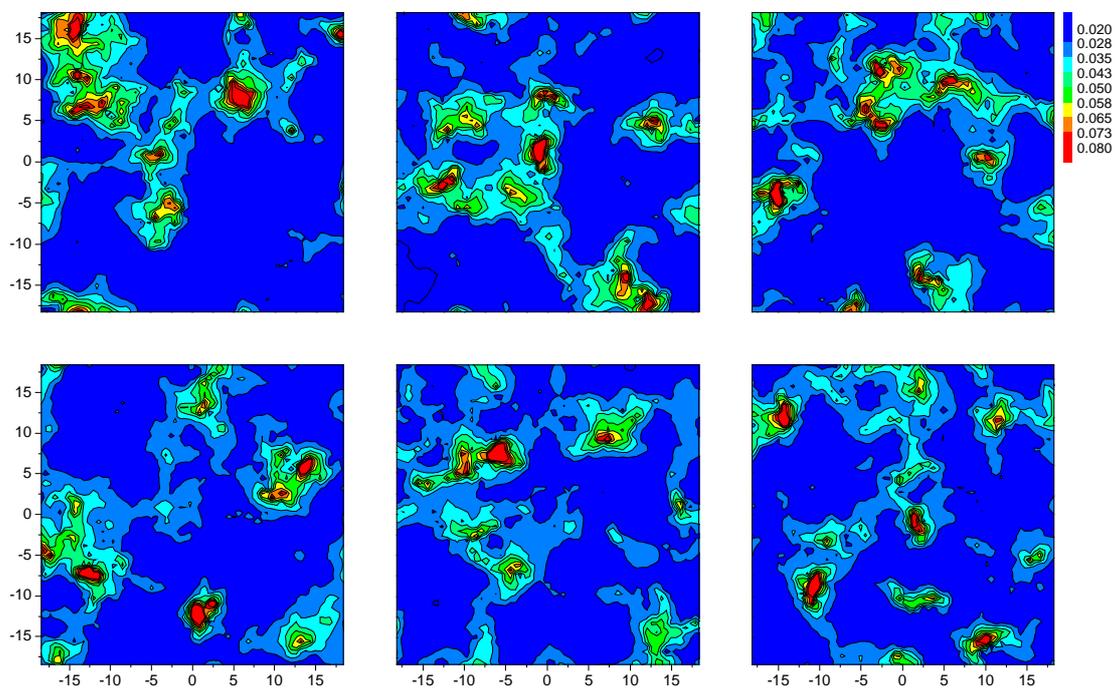

Figure 2

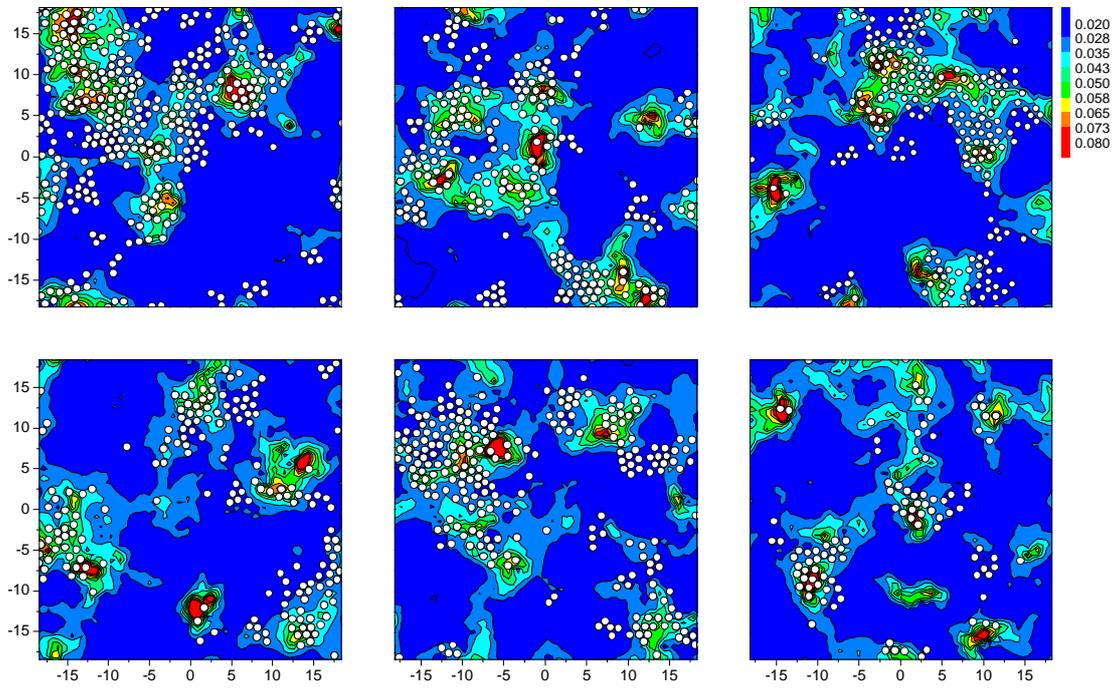

Figure 3

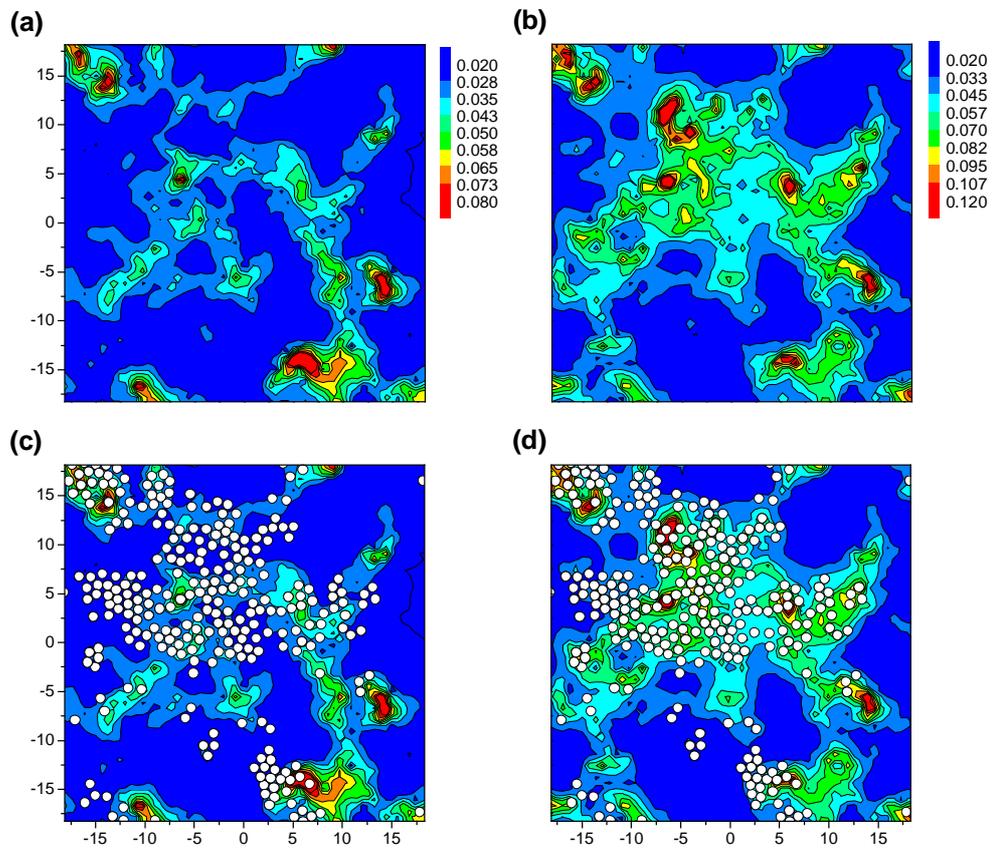

Figure 4

**Supplementary Information**

**1. Establishing the Irreversibility of Local Reorganization**

The probability that a particle could lose $x$ of its original neighbours and subsequently *fail* to recover $n$ of those lost neighbours was determined as follows. A hundred runs were carried out from a given starting configuration with momenta assigned randomly from an equilibrium Boltzmann distribution at the appropriate temperature. Run intervals of $1000\tau$ and $2000\tau$ were used. The initial neighbours of each particle were determined by using a cut-off distance equal to the distance to the first minimum in the species-appropriate radial distribution function.

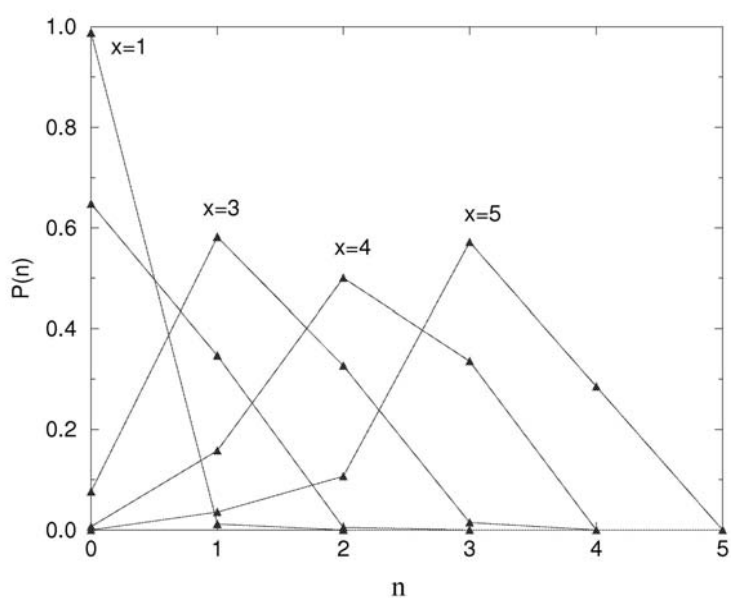

**Figure S1** The probability distributions for the minimum number of 'lost' initial neighbours $n$ after having first lost $x$ neighbours. Note that after losing $x = 4$ neighbours, the probability of recovering all of the initial neighbours (i.e. having n = 0) is essentially zero. The distributions show peaks at $n = x-2$ suggesting that particles typically recover 2 of their initial neighbours. These calculations where carried out over a time interval of $1000\tau$ at T = 0.4.

During the course of each trajectory, the neighbours of each particle were monitored. When particle $i$ first lost $x$ of its initial neighbours (with $x$ going from 1 to 5), the minimum number $n$ of *lost neighbours* that *failed* to be recovered in the remainder of the trajectory of particle $i$ was recorded. Complete recovery of the original neighbourhood corresponds to $n = 0$ lost neighbours that fail to be recovered. Note that a particle would be 'marked' as having $n = 0$, for example, even if it only recovered those neighbours for a very short time interval during the subsequent trajectory. Collecting data from each particle for all 100 runs, we plot the fraction of total observations corresponding to each value of $n$ in Fig. S1.

Losing a lot of bonds (i.e. large *x*) would be expected to take a longer time than losing a few bonds. That would mean that there would be less time left in the trajectory for the large *x* particles to recover their original neighbours. To see how serious this effect is, we repeated the calculations for a time interval of $2000\tau$, twice that used in Fig. S1. We found that the distributions shown in Fig. S1 were shifted to the right (i.e. towards higher *n*). We infer that increasing the run time will not alter our conclusion that the loss of 4 initial neighbours constitutes irreversible reorganization.

In the text we report on those particles that lost 4 original neighbours within $200\tau$. As we demonstrate the correlation between the spatial distribution of this irreversible relaxation and the low frequency normal modes of the initial configuration, it is of interest to know how the irreversible relaxation, as mapped in Figures 1, 3 and 4 evolves over the $200\,\tau$ time interval. In Figure S2 we have plotted the distribution of times at which particles first lost their 4$^{th}$ nearest neighbour. We clearly see that the time at which this condition for irreversibility is first being met is strongly weighted to the longer times.

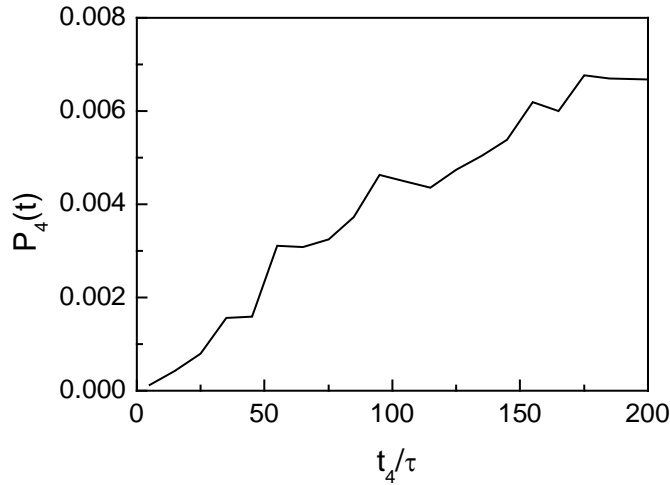

**Figure S2.** Distribution of times at which particles first lose their 4$^{th}$ nearest neighbour (data averaged over isoconfigurational ensembles of 100 runs for a total of 10 initial configurations of N=1024 particles at T=0.4).

### 2. Quasi-localised low-frequency normal modes

The localization of each mode is measured by the participation ratio,

$P(\omega) = \left[ N \sum_{i=1}^{N} (\vec{e}_\omega^i \cdot \vec{e}_\omega^i)^2 \right]^{-1}$. If the mode is completely delocalised so all particles contribute equally then $P(\omega) = 1$. At the other extreme, a mode localized on a single particle has $P(\omega) = 1/N$. For a plane wave, $P(\omega) = 2/3$.

In Figure S3 maps of the participation fractions for the lowest frequency modes for a single initial configuration are presented along with the corresponding irreversibility map. Quasi-localized modes are defined here as modes with a participation ratio < 0.34 and are coloured red. Heterogeneities are also observed in some delocalized (blue) modes. In all cases, these heterogeneous features of an eigenvector typically consist of a compact group of several particles that contribute significantly to the mode surrounded by an extended 'apron' of particles with smaller contributions. There is a visually striking correlation between regions of motion in the quasi-localized modes and regions of high probability of irreversible reorganisation.

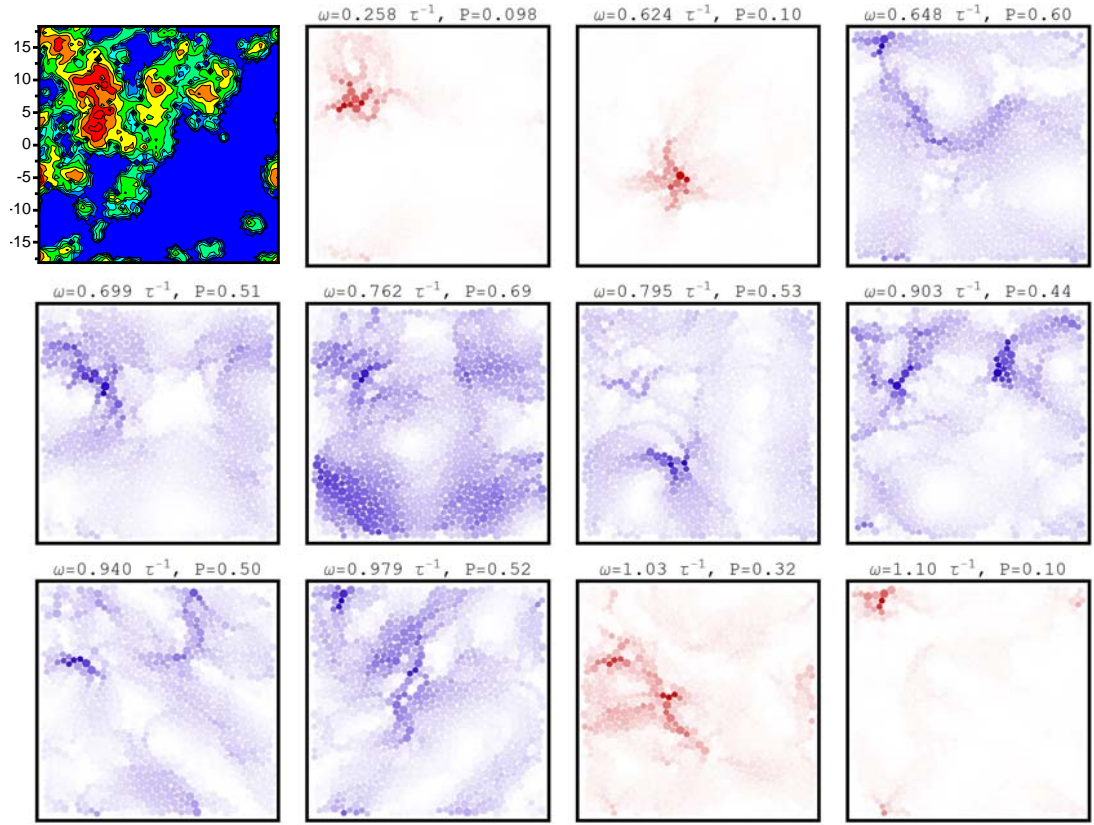

**Figure S3**. Maps of the participation fractions for the 11 lowest frequency normal modes of the local potential energy minimum associated with a single initial configuration. Modes with a participation ratio P < 0.34 are coloured red while the more delocalized modes are coloured blue. The intensity of the colour increases with the magnitude of the squared amplitude. The corresponding irreversibility map for this configuration is provided for comparison.

**3. Short time evolution of the participation fraction maps.**

In Figure S4 we plot the contour plots of the participation fraction summed over the 30 lowest frequency modes of the quenched configurations generated every $1\tau$ along a $10\tau$ trajectory. We remind the reader that $1\tau$ corresponds to ~10 collision times. We observe that the participation fraction maps undergo clear changes, indicative of a change in the local quenched potential minimum (or 'inherent structure'), while the overall spatial distribution of local low frequency modes changes little.

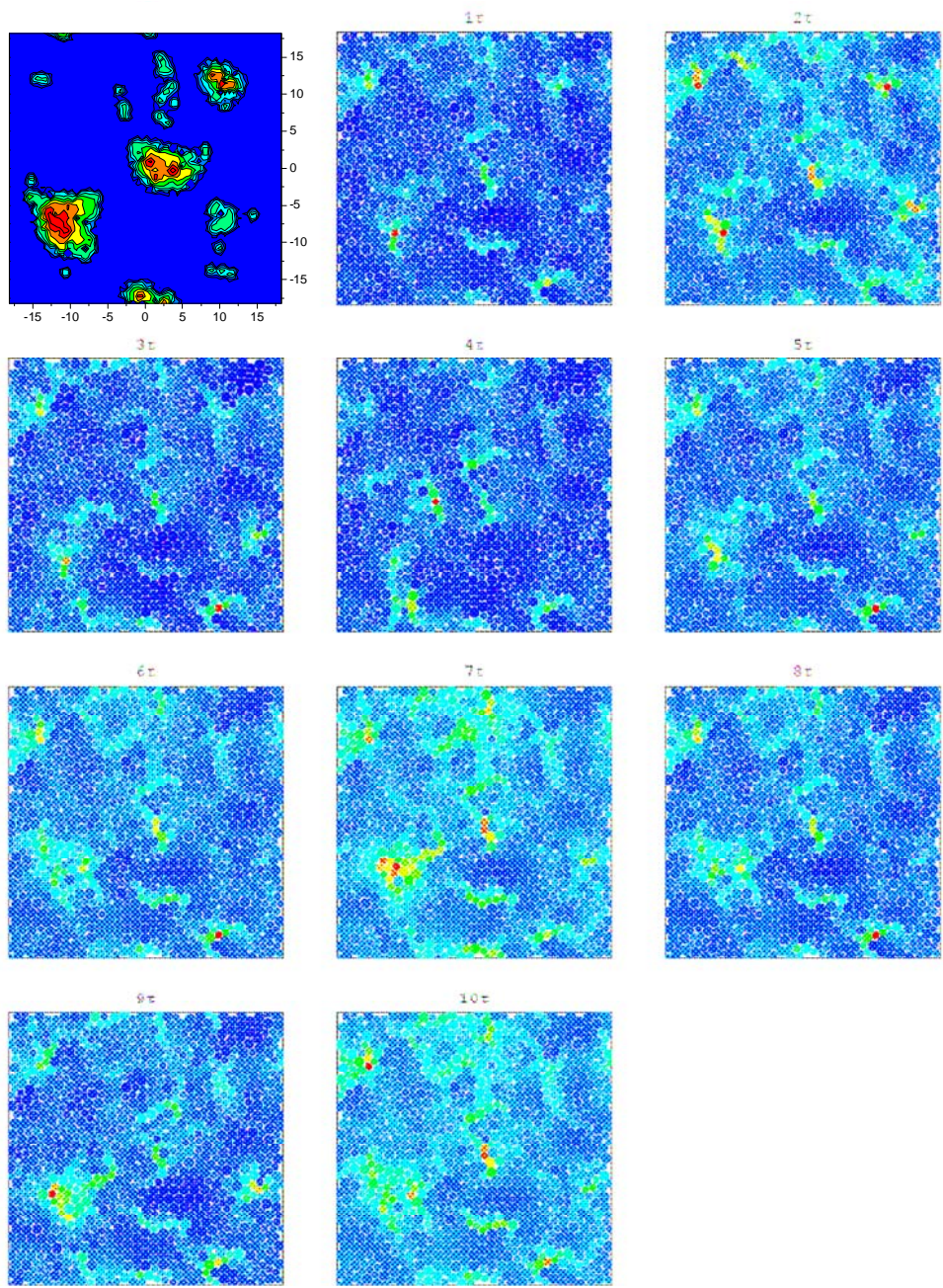

**Figure S4.** Plots of the participation fraction for quenched configurations taken every $1\tau$ along a $10\ \tau$ trajectory from the initial configuration. While there are clearly variations occurring in the distribution of modes (and hence in the identity of the quenched minimum or inherent structure) over $1\tau$, substantial elements of the mode distribution persist.

Unlike the configuration shown in Figure 4, many configurations show little significant difference between the low frequency participation map generated from the initial configuration and the map obtained by recording the maximum value of the participation fraction of each particle achieved over five 10τ runs. An example (the same configuration used in Fig.S4) is shown in Figure S5.

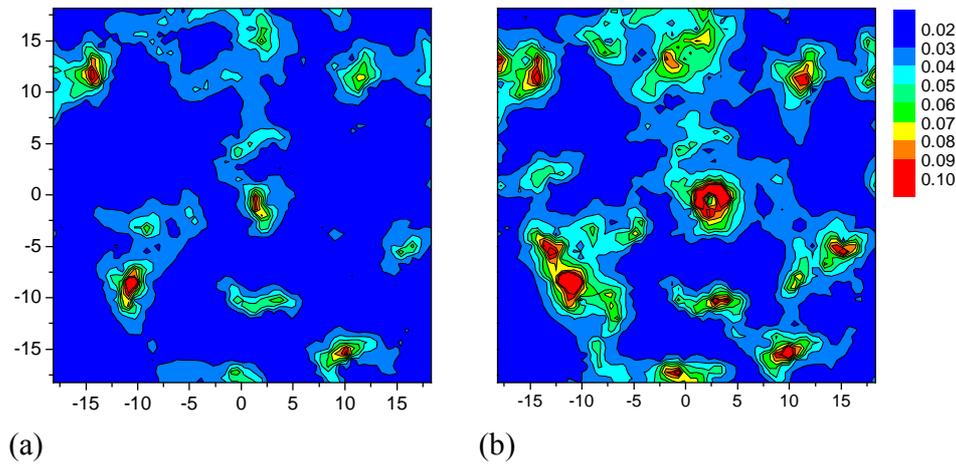

(a)                                  (b)

**Figure S5**. a) A map of the participation fraction of the quenched normal modes summed over the 30 lowest frequency modes. b) The map of the maximum value of the participation fraction achieved in five 10τ runs all starting from the same configuration analysed in 4a.

## 4. On the Similarity of Structure-Dynamics Correlations in 2D and 3D.

There is a striking correlation between the regions of motion in the quasi-localized modes ($P_\omega \leq 0.3$) and the regions of high Debye-Waller factors [A.Widmer-Cooper, P.Harrowell, *Phys. Rev. Lett.* **96**, 185701 (2006)] in both 2D and 3D. To quantify this, we look at particles with the top 10% highest values of the local Debye-Waller factor compared to the top 10% with the highest low-frequency motion. The low-frequency motion of each particle is obtained by summing its participation fraction, $p_i$, for all low frequency modes. (As the number of low frequency plane wave modes increases with system size, some upper limit in participation ratio may need to be applied for larger systems.) Summing between 30-80 of the lowest frequency modes gives qualitatively similar results. The overlap of the regions of large short-time displacement large normal mode displacement are computed by first eliminating particles from both maps that are not in a cluster of at least n* = 3 to eliminate noise and focus on regions of collective motion. The percentage of particles in clusters of the top 10% for both the local Debye-Waller factor and the summed 30 lowest frequency modes for 10 different configurations are plotted in Fig. S6, along with the overlap of clusters of a random sample of 10% of the particles for comparison.

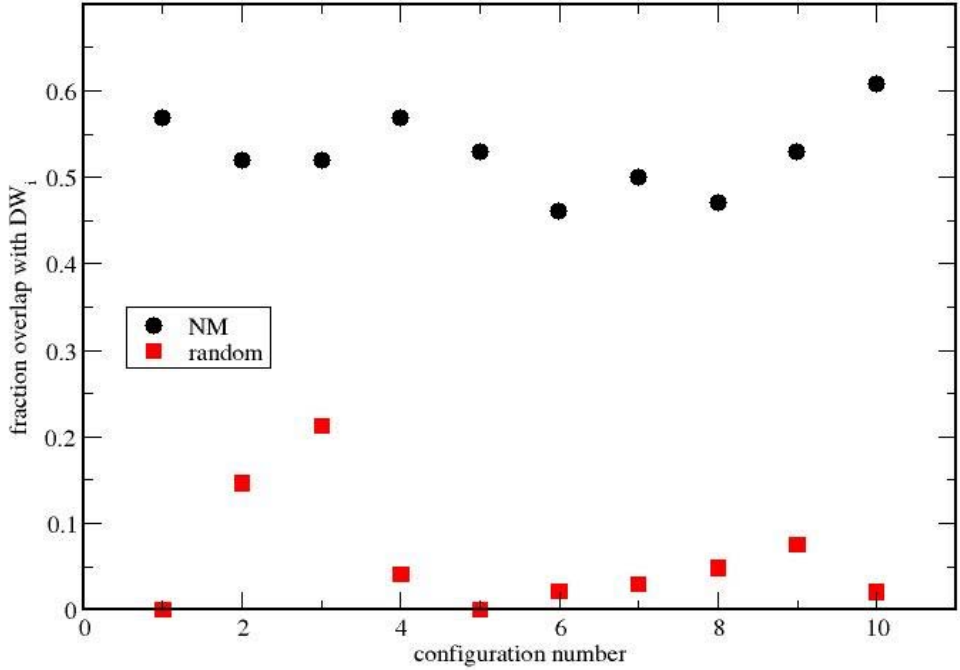

**Figure S6**. For the 2D system, the overlap of the 10% of particles with the largest Debye-Waller factors with (black dots) the top 10% of clustered particles with the largest amplitudes in the sum of the lowest 30 normal modes and with (red squares) a random selection of 10% of the particles, all maps have had clusters of less than n*=3 eliminated.

To insure that this correlation holds in 3D, we performed a similar analysis on a 3D binary soft-sphere system. These particles interact with a pairwise repulsion,

$$u_{ab}(r) = \varepsilon \left[ \frac{\sigma_{ab}}{r} \right]^{12} \tag{S1}$$

with $\sigma_{12}$ = 1.1 x $\sigma_{11}$, $\sigma_{22}$ = 1.2 x $\sigma_{11}$ and all units reduced such that $\sigma_{11}$ = 1.0, $\varepsilon$ = 1.0, m = 1.0 and the time unit $\tau = \sigma_{11}(m/\varepsilon)^{1/2}$. Equilibrated configurations were obtained from P. Verrocchio [L. A. Fernández, V. Martín-Mayor, P. Verrocchio, *Phys. Rev. E* **73**, 020501(R) (2006)] for temperature down to $0.92T_{MC}$, where the mode-coupling temperature [W. Götze, L. Sjögren, *Rep. Prog. Phys*. **55**, 241 (1992)] is $T_{MC} = 0.2084$ $k_B/\varepsilon$ [T. S. Grigera, G. Parisi, *Phys. Rev. E*. **63** 045102 (2001)]. A microcanonical molecular dynamics simulation was carried out for N = 1024 particles (50% each small and large)) at a density $\rho$ = 0.7421 $\sigma_{11}^{-3}$ and temperature $T = 0.2084$ $k_B/\varepsilon$ in a cube box with periodic boundary conditions. The alpha relaxation time was found to be $\tau_\alpha$ = 832 $\tau$ and the non-Gaussian parameter peak at 200 τ. The local Debye-Waller factor for each particle was calculated as the variance of 100 runs of the iso-configurational distribution of each particle's displacement over a time interval of 10 τ. The same normal-mode analysis was performed on this system as on the 2D system, but the dimensionality of the eigenspace is 50% larger than for the 2D system. In 3D, analyzing the overlap of the clusters of the 10% of particles with the largest local Debye-Waller factor and those with the highest summed amplitude from the low-frequency modes show that anywhere between 45-150 modes gives a good prediction of the short-time dynamics. Clusters in 3D are defined with a minimum size of n*= 5. Fig. S7 shows the overlap between the most mobile particles at 10 τ and the summed amplitudes of the 60 lowest frequency modes.

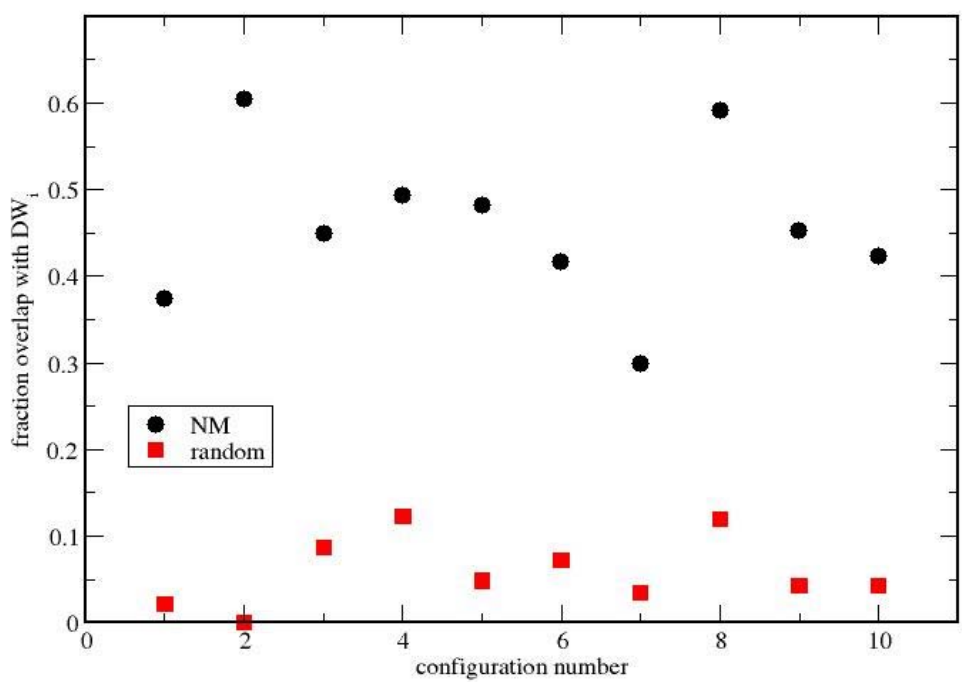

**Figure S7.** Similar to Figure S6 for the 3D system, in this dimension the minimum cluster size is n*=5.

**5. Correlation Between IR Particles and Mode Maps.**
Here we perform an analysis of the overlap between the IR particles and the mode maps in a manner similar to that described above. We first define a cluster of particles if 5 or more particles form a domain for which there is at least one nearest neighbour particle that belongs to the domain for each particle that lies in the domain. We do this for both the particles in the mode map that lie in the "non-blue" areas (those above a threshold of 0.028 in the units of Fig.3 of the text) as well as for the IR particles. We define "overlap" as the % fraction of IR particles in clusters that coincide with the particles that exist in the clusters of the mode map. We then measure the overlap of the maps for each of the 10 independent configurations we have used. Then, for each configuration we generate a map at random that contains the same number of particles as in the IR map for that given configuration. Using the same algorithm, we also compare the resulting overlap with the particles in the mode-map. This is plotted in Fig.S8 below.

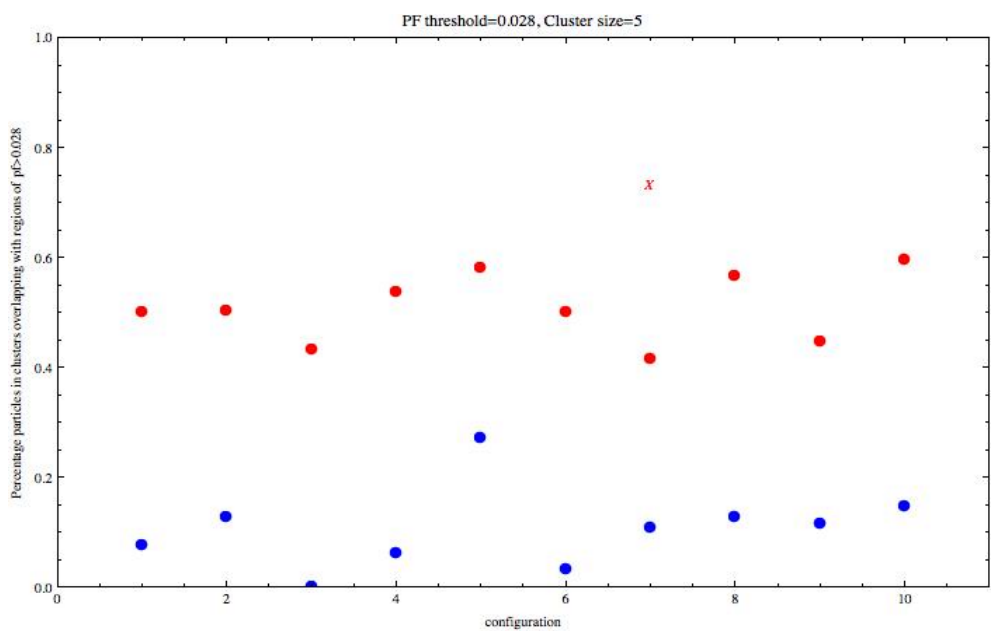

**Figure S8.** Comparison of overlap of IR particle clusters to mode clusters (red) and random clusters (blue). Overlap of IR clusters and mode clusters (as defined above) is, on average, about 50%, with only 3 of ten configurations having an overlap in the range of 40%. For the configuration with the smallest overlap (the configuration presented in Fig.4 of the original text) the modes that participate in an initial short run (as defined in the original text) produces a "renormalized mode map" with a cluster overlap close to 80% as indicated by the "x"

## 6. Finite Size.

Here we investigate one configuration of our system prepared at the same temperature but for 4096 particles. As for the smaller system studied in the main text, we compare the IR particles to the mode map in Fig.S9. For this particular configuration, the overlap as defined above is 68%. The conclusion drawn from this is that there are not significant finite size effects that alter the conclusions drawn in the text.

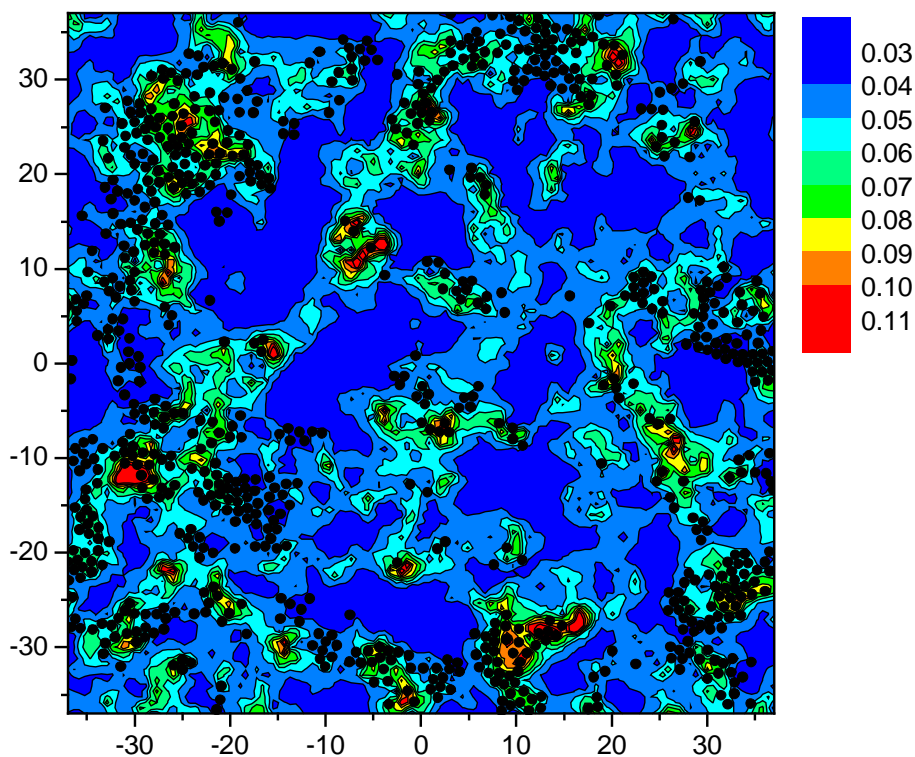

**Figure S9.** Contour plots for an equimolar mixture of 4096 particles of the low frequency mode participation (as in Fig. 2 of the paper), overlaid with the location of particles (filled circles) whose iso-configurational probability of losing 4 initial nearest neighbours within $200\tau$ is greater than or equal to 0.01.

## 7. On The Construction of Mode Maps.

As mentioned above, the results presented in the text are not sensitive to how many modes are used to construct the mode maps, up to a point. If we plot the participation ratio vs. frequency, we find that in the low frequency band the participation ratio (averaged over small frequency bins containing a few modes each) drops from large values at the lowest frequencies, then has a dip and then flattens out to a relatively constant value. At higher frequencies still the participation ratio decreases sharply (reflecting strict localization). As long as the cutoff on the number of modes is defined so as not to extend *deeply* into the intermediate (flat) or localized high frequency band, then the results are very not sensitive (as long as enough modes are used, of course). If the cutoff extends past the center of the intermediate frequency regime where the participation ratio is relatively constant, the results degrade, but not quickly. This simply reflects the fact that the higher frequency (truly localized) modes do not contribute to the spatial properties of IR regions. In this sense the results are not sensitive, as long as a (well defined) frequency band is used. Results are similar if only low-participation modes are used, but since the definition of "low" is not so clear, we have decided to take an unbiased sum where the localized regions naturally stand out in relief with regard to the mixing of any plane wave character that is carried along.